\documentclass{emulateapj}

\usepackage{hyperref}

\begin{document}


\title{Detection of a Nearby Halo Debris Stream in the WISE and 2MASS Surveys}

\author{Carl J. Grillmair}
\affil{Spitzer Science Center, 1200 E. California Blvd., Pasadena, CA 91125, USA}

\author{Roc Cutri, Frank J. Masci, Tim Conrow}
\affil{Infrared Processing and Analysis Center, California Institute of Technology, Pasadena, CA 91125}

\author{Branimir Sesar}
\affil{Division of Physics, Mathematics, and Astronomy, California Institute of Technology, Pasadena, CA 91125}

\author{Peter R. M. Eisenhardt}
\affil{Jet Propulsion Laboratory, California Institute of Technology, MS 169-327, 4800 Oak Grove Drive, Pasadena, CA 91109}

\author{Edward L. Wright}
\affil{Department of Physics, University of California, Los Angeles, CA 90095}

\begin{abstract}

Combining the Wide-Field Infrared Survey Explorer All-Sky Release with
the 2MASS Point Source Catalog, we detect a nearby, moderately
metal-poor stellar debris stream spanning $24\arcdeg$ across the
southern sky.  The stream, which we designate Alpheus, is at an
estimated distance of $\sim 1.9$ kpc. Its position, orientation,
width, estimated metallicity, and to some extent its distance, are in
approximate agreement with what one might expect of the leading tidal
tail of the southern globular cluster NGC 288.

\end{abstract}


\keywords{globular clusters: general --- globular clusters: individual (NGC 288) --- Galaxy: Structure --- Galaxy: Halo}

\section{Introduction}

Wide-field photometric surveys have enabled the identification of at
least 15 stellar debris streams in the Galactic halo
\citep{grill2010,bonaca2012,martin2013}. Roughly the same number of
cold substructures have been found in velocity space
\citep{helmi1999,smith2009,schlaufman2009, williams2011}.  In addition
to helping us reconstruct the accretion history of the Galaxy, they
can also serve as sensitive probes of the Galactic potential
\citep{law2009, koposov2010}. Globular cluster streams are
particularly interesting in this regard as they are the coldest
stellar substructures yet discovered \citep{combes1999,
  odenkirchen2009, willett2009}. A large sample of such streams will
ultimately enable us to map the distribution of Galactic dark
matter with much greater spatial resolution than has been possible to
date. Detailed analyses of these streams may also enable us to better
characterize the dark matter sub-halos believed to populate the
Galactic halo.  \citep{yoon2010, carlberg2009, carlberg2012}.

In this paper we examine the 2MASS Point Source Catalog
\citep{skrutskie2006} and the 
Wide Field Infrared Survey Explorer (WISE) All-Sky Release \citep{wright2010} for
evidence of nearby streams in regions of the sky as yet unexplored in
other surveys.  We briefly describe our analysis in Section
\ref{analysis}. We characterize the stream in Section \ref{discussion}
and we discuss evidence that the stream may be associated with the
globular cluster NGC 288 in Section \ref{ngc288}. We make concluding
remarks Section \ref{conclusion}.

\section{Data Analysis} \label{analysis}

Data comprising WISE bands W1 (3.4\micron) and W2 (4.6\micron), along
with matching 2MASS J (1.25\micron) photometry, were extracted from
the WISE All-Sky Release and analyzed using the matched filter
technique described by \citet{rock2002} and \citet{grill2009}.
Matched-filtering has been used to detect halo streams to very low
surface densities in Sloan Digital Sky Survey data
\citep{grillj2006,grilld2006a,grilld2006b,grill2009,grill2011,bonaca2012}. While
2MASS data have been used to map the Sagittarius stream using M giants
\citep{maje2003}, both WISE and 2MASS data are more limited in this
application for main sequence stars due to their much fainter
magnitudes. SDSS experience has shown that matched-filter techniques
over large areas work well only if the main sequence turnoff of a
population is detectable. Moreover, while WISE goes somewhat deeper
than 2MASS J, the W1-W2 colors of stars are very nearly degenerate
compared with the photometric uncertainties. Combining WISE bands W1
and W2 with 2MASS J improves the situation, providing both color
discrimination and greater depth than is possible with 2MASS
photometry alone. With a limiting magnitude of J $\approx 16.5$, the
WISE and 2MASS data can probe main sequence populations out to a
distance of $\approx 4$ kpc.

An additional issue with the WISE All-Sky data are the presence of
significant, magnitude-dependent, photometric biases at faint
magnitudes resulting from an overestimate of background
levels\footnote{\url{http://wise2.ipac.caltech.edu/docs/release/allsky/expsup/sec6_3c.html}}.
Since these biases can exceed 30\% at the faintest magnitudes, this
has serious consequences for filtering of sources based on
color. Aperture magnitudes are less affected by this bias but they are
much noisier due to aperture contamination. We consequently measured
this bias in W1 and W2 by comparing profile-fitting measurements with
aperture photometry using several million well-fit, isolated stars
in the region $0\arcdeg < $ R.A. $ < 45\arcdeg$, $-60\arcdeg < $
dec $ < -30\arcdeg$. We generated a look-up table of
photometric bias versus magnitude at half-magnitude intervals and
interpolated this table to compute appropriate corrections for
profile-fitted photometry.

We constructed matched filters based on the Padova database of
theoretical stellar isochrones \citep{marigo2008, girardi2010}.  All
stars with $0 < J < 16.5$ were used, and we dereddened both WISE and
2MASS photometry as a function of position on the sky using the
DIRBE/IRAS dust maps of \citet{schleg98}, corrected using the
prescription of \citet{schlafly2011}.  We use only stars with CC flags
of 0, photometric quality flags of ``A'', ``B'', or ``C'', $\chi^2 <
3.0$, and the ext flag = 0, to select on uncontaminated, unresolved
objects with SNR $> 5$.  For the present search, the color-magnitude
distribution of field stars was sampled using $\sim 1.2 \times 10^7$
stars in a broad region extending from the Small Magellanic Cloud
(SMC) to the south Galactic pole. The filters were simultaneously
applied to the entire WISE and 2MASS catalogs, and the resulting
weights were combined to yield all-sky, filtered surface density maps.

Figure 1 shows the filtered star count distribution in the region of
the south Galactic pole using a filter whose basis is an isochrone
with Z = 0.0025, an age of 12 Gyrs, and is shifted so as to optimally
sample populations at a distance of 1.9 kpc.  This distribution was
constructed by combining the weights of stars in 0.2\arcdeg~ bins
based on their distances from the $J - W1$ and $J - W2$ main sequence
and giant branch color-magnitude loci.

\section{Discussion} \label{discussion}

Apparent in Figure 1 is a long, narrow feature spanning some
24\arcdeg~ across the southern sky. The orientation of the feature
does not precisely align with lines of either constant R.A. or dec, or
constant ecliptic longitude or latitude, indicating that the feature
is not an artifact of sampling or data quality discontinuities in
either of the two surveys. The feature shows no correlation with
reddening maps of the area (where E(B-V) is typically $\sim 0.02$),
nor does it correlate with the 2MASS j\_msnr10 values\footnote{\url{
    http://www.ipac.caltech.edu/2mass/releases/allsky/doc/sec6_2.html}},
which estimate the depth of the 2MASS survey as a function of sky
position. The feature is also localized in color-magnitude space,
which would not be the case if the feature were due simply to
variations in completeness. We conclude that the feature is not an
artifact, but a bona fide stellar debris stream. Other curvilinear
features are evident at various locations across the sky, but at much
lower signal-to-noise ratios.

The visible stream runs through Phoenix, Eridanus, Hydrus, Tucana, and
becomes lost at its southern end near the SMC. Given the rather blue
filters we are using and the magnitude limits of the WISE and 2MASS
surveys, we regard it as extremely unlikely that the new stream could
be associated with the SMC.  Moreover, the orientation of the stream
is nearly perpendicular to the SMC's orbital path
\citep{piatek2008}. We designate the new stream Alpheus, after the
river in the {\it Illiad.} The northern end of the stream
appears to end fairly abruptly at (RA, dec) = (27.7\arcdeg,
-45.0\arcdeg).  Over the interval $-69 < \delta < -45$, the path of
the stream is well fit in celestial coordinates (to within 1\arcdeg)
by a polynomial of the form:

\begin{equation}
\label{trace}
\alpha = 32.116 - 0.00256 \times \delta - 0.00225 \times \delta^2 
\end{equation}

Color-magnitude Hess diagrams (CMDs) for Alpheus are shown in
Figure 2. These distributions were determined using Equation 1 to
select stars within $1\arcdeg$ of the centerline of the
stream. Similar regions, $4\arcdeg$ wide on either side of the stream
and laterally offset by $5\arcdeg$, were used to sample the field star
population. Scaling the latter to the former by area, the CMDs in
Figure 2 are subtractions of the two. While not strong, a turn-off and
upper main sequence are clearly visible. For J - W1, the sequence is
matched reasonably well using a Padova isochrone having [Fe/H] = -1.0,
though the uncertainty is quite large because the isochrones for
different metallicities are rather closely spaced at these
wavelengths. The J - W2 CMD is less convincing, suggesting a somewhat
bluer sequence.  Based on the stronger J - W1 sequence, Figure 2
suggests an uncertainty on the order of at least 0.5 dex.  Deeper
follow-up surveys in visual bands and spectroscopy of individual stars
will be required to better constrain the metallicity of the stream.

A number of curvilinear features are evident in Figure 1, and one
might ask how Alpheus compares in terms of significance. We can make
use of the ``T-statistic'' of \citet{grill2009}, which measures the
median contrast along its length between a putative stream and the
surrounding field. The T-statistic for Alpheus, comparing with the
field extending the length of the stream and $30\arcdeg$ to the north
and $13\arcdeg$ to the south in ecliptic coordinates, is shown in
Figure 3. The stream is detected at the $19\sigma$ level, and the
lateral full-width-at-half-maximum of the stream is $3.2\arcdeg$. At a
distance of 1.9 kpc (see below) this corresponds to a spatial extent
of $\approx 100$ pc. This is similar to the widths of other streams
that are known or believed to have been generated by globular clusters
\citep{odenkirchen2003, grillj2006,grilld2006b, grill2009, grill2011},
suggesting that the progenitor of Alpheus was itself a globular
cluster. The next-most convincing overdensity is detected at $<
10\sigma$, and we will consider this and other features in a
forthcoming paper.

Following \citet{grilld2006b} we shift our filter brightward and
faintward to estimate the stream's distance. Combining results for J -
W1 and J - W2, we find that the signal-to-noise ratio of the northern
half of the stream peaks at a distance modulus of $11.5 \pm 0.4$ mag,
while the southern half of the stream peaks at $11.0 \pm 0.5$ mag. The
uncertainties are quite large due to a combination of the small number
of stars, large photometric errors at the faint limit, and the
relatively narrow range of stellar colors at these wavelengths. This
would put the northern half of the stream at a mean sun-centric
distance of $2.0 \pm 0.7$ kpc, while the southern half is at $1.6 \pm
0.8$ kpc. We note that a simple ratio of filtered stream to
background signal peaks at 2.3 kpc in the southern half and 3.9 kpc in
the northern half. However, the stream is not nearly as apparent or
continuous using such a filter, nor does such a filter match the CMD
in Figure 2. 

Summing the color-selected, background-subtracted, but otherwise
unweighted star counts over a width of $3.2\arcdeg$ we find the number
of stars in the stream to W1 $\approx 16.5$ to be $310 \pm 90$.  The
quoted uncertainty is largely due to the shot noise contributed by the
several thousand field stars with similar colors lying along our line
of sight to the stream. The average surface density is $4 \pm 1$ stars
deg$^{-2}$, with peaks of $\sim 10$ stars deg$^{-2}$.

\subsection{On the Possible Association of Alpheus with the Globular Cluster NGC 288}\label{ngc288}

Based on its relatively low concentration and fairly eccentric,
retrograde orbit, NGC 288 has a relatively high predicted destruction
rate \citep{gnedin1997}, and both \citet{grill1995} and
\citet{leon2000} found a significant population of extratidal stars in
the vicinity of the cluster.  Figure 4 shows projections of three
possible orbits of the cluster, integrated using the Galactic model of
\citet{allen91}. All integrations use a cluster distance of 8.3 kpc
and a radial velocity of -47 kms$^{-1}$ \citep{harris1996}. The three
different orbits are based on proper motions of
$\mu_{\alpha}\cos{\delta} = 4.68 \pm 0.2 $ mas/yr, $\mu_{\delta} =
-5.25 \pm 0.2$ mas/yr \citep{guo1995} ('a'), $\mu_{\alpha}\cos{\delta}
= 4.64 \pm 0.4$ mas/yr, $\mu_{\delta} = -6.00 \pm 0.4$ mas/yr
\citep{dinescu1997} ('b'), and $\mu_{\alpha}\cos{\delta} = 3.8 \pm
0.3$ mas/yr, $\mu_{\delta} = -8.1 \pm 0.3$ mas/yr \citep{platais1998}
('c').

Two of the orbit integrations ('a' and 'b') lie nearly parallel to the
new stream, offset to the east by $\sim 2-3\arcdeg$. Orbit 'c' is also
nearly parallel to the stream, but is offset $\sim 7\arcdeg$ to the
west. Tracing around the integrated orbits, the visible portion of the
stream begins at $\sim 4$ kpc from the cluster and ends at $\sim 6.3$
kpc. Over the 24\arcdeg~ extent of the stream, the leading arm of orbit
'a' predicts a heliocentric distance ranging from 3.8 kpc at the
southern end to 5.1 kpc at the northern end. Similarly, orbit 'b'
ranges from 5 kpc to 6.2 kpc, and orbit 'c' ranges from 7.2 kpc to 7.5
kpc. The near coincidences in the position, orientation, and distance
gradient of the stream, particularly with orbit 'a', suggest a
possible association with NGC 288.

Given the retrograde motion of NGC 288, the stream would constitute
the leading arm of the NGC 288's tidal tails. We would expect these
stars to have dropped inwards towards the Galactic center before being
carried forward along their new orbits. That the stream lies to the
inside of putative orbits 'a' and 'b' is consistent with this
picture. An offset of 2.5\arcdeg~ at a distance of 1.9 kpc corresponds to
$\approx 80$ pc in projection. Though this offset is somewhat larger
than the current tidal radius of NGC 288 ($\sim 31$ pc,
\citet{harris1996}), we expect escaped stars to have residual
velocities that will carry them somewhat farther from the
cluster.  Indeed, \citet{grill1995} and \citet{leon2000} found
extratidal stars extending to at least 150 pc in the immediate
vicinity of NGC 288.

Our metallicity estimate for stars in the stream is also consistent
with a measured value of [Fe/H] = -1.24 (Z $\approx 0.0015$) for NGC
288 \citep{harris1996}. As noted earlier, while the stream appears
strongest using isochrones with Z = 0.0025 (Fe/H $\approx -1.0$), due
to the small number of stars and the narrow range of colors at these
wavelengths, the stream is only marginally less apparent (S/N $\sim
17.5$) using an isochrone with Z = 0.0015. 2MASS and WISE colors are
not very sensitive to metallicity, and given the additional
uncertainty contributed by our substantial corrections to the WISE
profile-fit photometry, we consider the agreement to be quite good.

If we adopt a proper motion for NGC 288 of $\mu_{\alpha}\cos{\delta} =
4.75 $ mas/yr, $\mu_{\delta} = -4.60$ mas/yr, we arrive at orbit 'd'
in Figure 4. This value of the proper motion was arrived at by both
forcing the orbit to pass through NGC 288, and to pass 2.5\arcdeg~ to
the east along the length of Alpheus.The heliocentric distance of this
orbit in the range $-65\arcdeg < \delta < -45\arcdeg$ is reduced to
between 2.3 and 3.1 kpc which, to within the uncertainties, is in
agreement with our distance estimate for the stream. The adopted value
of $\mu_{\alpha}\cos{\delta}$ is in good agreement with the measured
values, while the value of $\mu_{\delta}$ is $1.2\sigma$ from the mean
of the three measurements.  Combining the measurement uncertainties
with inaccuracies in our adopted model for the Galactic potential, we
consider the association of the stream with NGC 288 plausible, if not
proved. Validation of our hypothesis will require radial velocity
measurements for a significant number of stream stars.

That the stream does not visibly connect with NGC 288 in Figure 4 may
simply be due to the limiting magnitude of the WISE and 2MASS
surveys. At the northernmost point of the visible stream, the NGC 288
orbit model predicts that the heliocentric distances rapidly increase
to 8.3 kpc at the position of NGC 288. As we use our
filters to probe beyond 3-5 kpc, the turn-off and subgiant branch
fall below our limiting magnitude and we are forced to rely wholly on
the RGB for our signal. However, the RGB is much more sparsely
populated and more highly contaminated by foreground stars than is
the turn-off region. We detect NGC 288 itself only
by virtue of the relatively large number of RGB stars in the
cluster itself.

It is also possible that the stream has been partially
disrupted by a dark matter subhalo or other perturber of significant
mass and/or favorable trajectory. However, examinations of the Pal 5
and GD-1 streams \citep{carlberg2012, carlberg2013} have not
identified gaps anywhere near 4 kpc in length. We consequently
consider this an unlikely possibility, but one which will need to be
tested with deeper observations.

\section{Conclusion} \label{conclusion}

Using the 2MASS and WISE All-Sky Catalogs, we identify a
new, 24\arcdeg-long, moderately metal-poor stellar debris stream in the
southern hemisphere which we designate Alpheus. At a distance of
$\approx 1.9$ kpc, the stream is much nearer than other spatially
mapped streams that have been found to date.  

Based on near coincidences in position, orientation, distance,
distance gradient, and metallicity, we suggest that the stream may be
the leading tidal arm of the globular cluster NGC 288. Testing this
hypothesis will require both deeper imaging in the visible bands and
spectroscopy. Deeper imaging by the SkyMapper survey
\citep{keller2007} should enable us to reach well down the main
sequence to better characterize the stream and extend its path across
the sky.  Based on existing proper motion measurements for NGC 288 as
well as our attempt to estimate an orbit that would connect Alpheus
with the cluster, we would predict radial velocities in the range $-66
< v_r < -36$ km s$^{-1}$ at the northern end of the stream, and $23 <
v_r < 36$ km s$^{-1}$ at the southern end. If indeed Alpheus is
associated with NGC 288, these measurements will also enable more
comprehensive modeling of the orbit of NGC 288 and the shape of the
Galactic potential towards the south Galactic pole.

\acknowledgments

We gratefully acknowledge several probing questions by an anonymous
referee which helped us to improve and clarify the manuscript.  We
also thank S. van den Bergh for suggesting a new source of stream
designations. This publication makes use of data products from the
Wide-Field Infrared Survey Explorer, which is a joint project of the
University of California, Los Angeles, and the Jet Propulsion
Laboratory/California Institute of Technology, funded by the National
Aeronautics and Space Administration. It also makes use of data
products from the Two Micron All Sky Survey, which is a joint project
of the University of Massachusetts and the Infrared Processing and
Analysis Center, funded by the National Aeronautics and Space
Administration and the National Science Foundation.

{\it Facilities:} \facility{WISE, 2MASS, IRSA}.

\clearpage

\begin{figure}
\epsscale{0.75}
\plotone{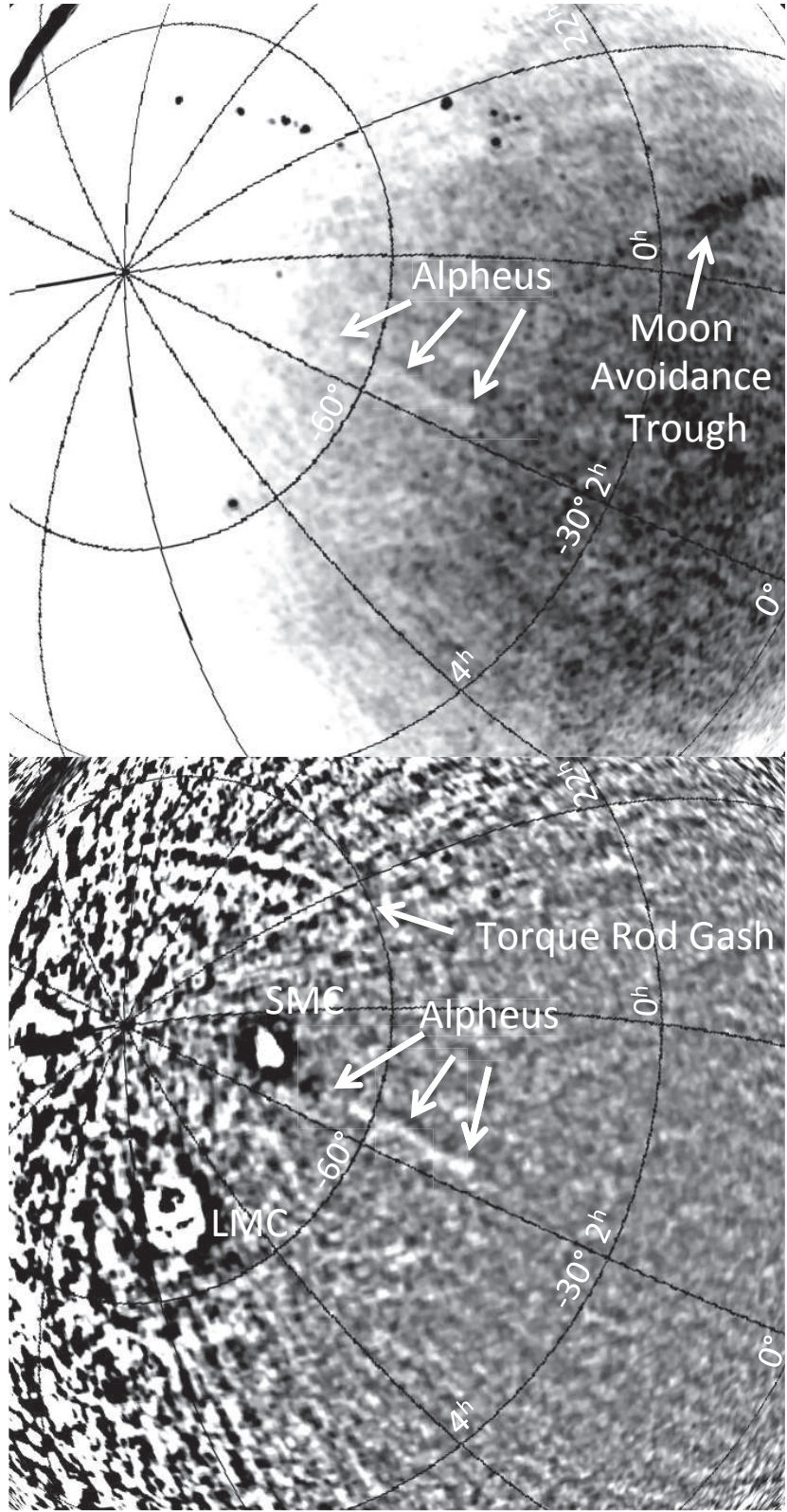}
\caption{High contrast, filtered surface density map of WISE/2MASS
  stars in the south Galactic cap.  The upper panel shows the map with
  a log stretch, smoothed with a Gaussian kernel of width
  $0.4\arcdeg$. Lighter areas indicating higher surface densities. The
  map is the result of a filter based on a Padova isochrone with
  [Fe/H] = -1.0, an age of 12 Gyr, and shifted to a distance of 1.9
  kpc. In the lower panel, the map has been background-subtracted
  using a window-smoothed version of itself (unsharp-masking,
  e.g. \citet{lee2009}) to remove power at low frequencies, reduce the
  dynamic range, and make various features and landmarks more apparent
  (e.g. the LMC and SMC).  The map is smoothed as above and is shown
  with a linear stretch.  WISE-specific features resulting from
  moon-avoidance maneuvers and momentum dumps (repeated firing of the
  spacecraft torque rods at nearly the same ecliptic latitude) are
  indicated. The curved structure between the LMC and SMC is part of a
  ring-like feature surrounded the south ecliptic pole and is an
  artifact of image ``decimation'' in the WISE photometry pipeline,
  required in high-coverage areas due to computational limitations.}

\end{figure}

\begin{figure}
\epsscale{1.0}
\plotone{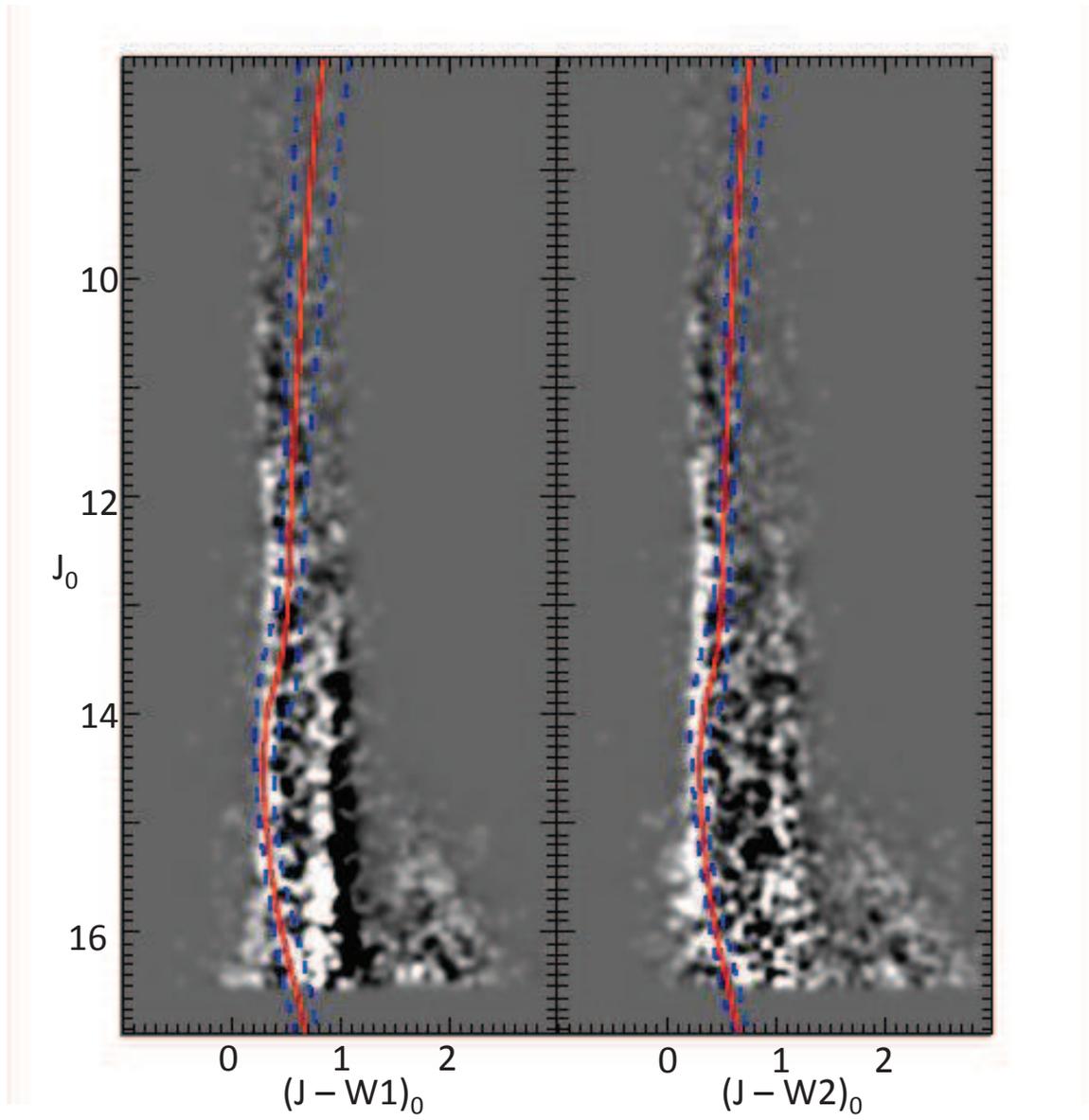}
\caption{Hess diagrams of stars lying within $1\arcdeg$ of the
  centerline of Alpheus. The solid line shows a Padova isochrone with
  [Fe/H] =-1.0, age 12 Gyrs, and shifted to a distance of 1.9 kpc. The
  dashed lines show similar isochrones for [Fe/H] = - 2.4 and solar
  metallicity. Lighter areas indicate higher surface densities.}

\end{figure}

\begin{figure}
\epsscale{1.0}
\plotone{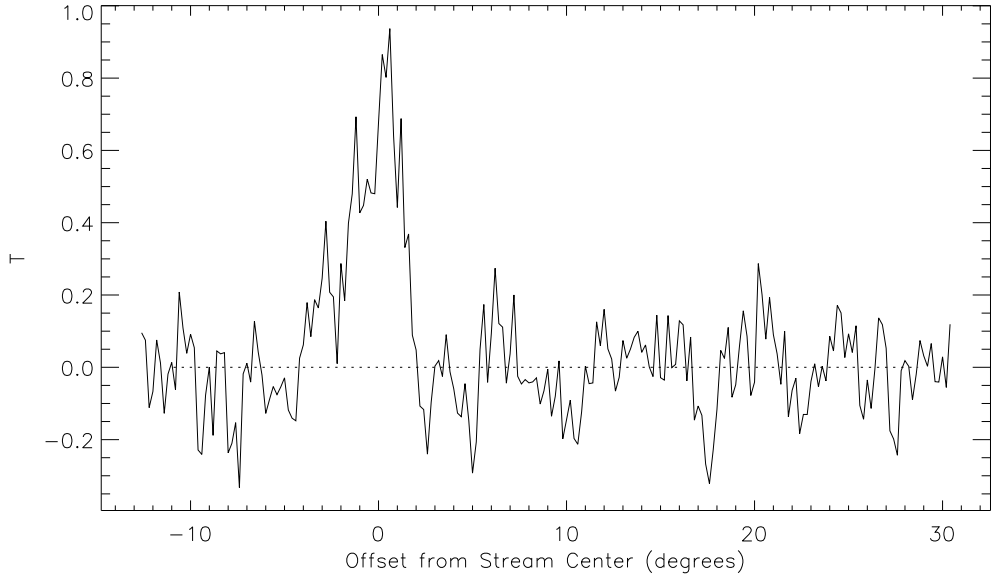}
\caption{The lateral profile of the stream, as measured using the T
  statistic (arbitrary units). The profile measures the median
  signal-to-noise ratio in five, 5\arcdeg~ segments of Alpheus, where
  the noise is measured in fields extending 13\arcdeg~ and 30\arcdeg~
  to the south and north, respectively.} 

\end{figure}

\begin{figure}
\epsscale{1.0}
\plotone{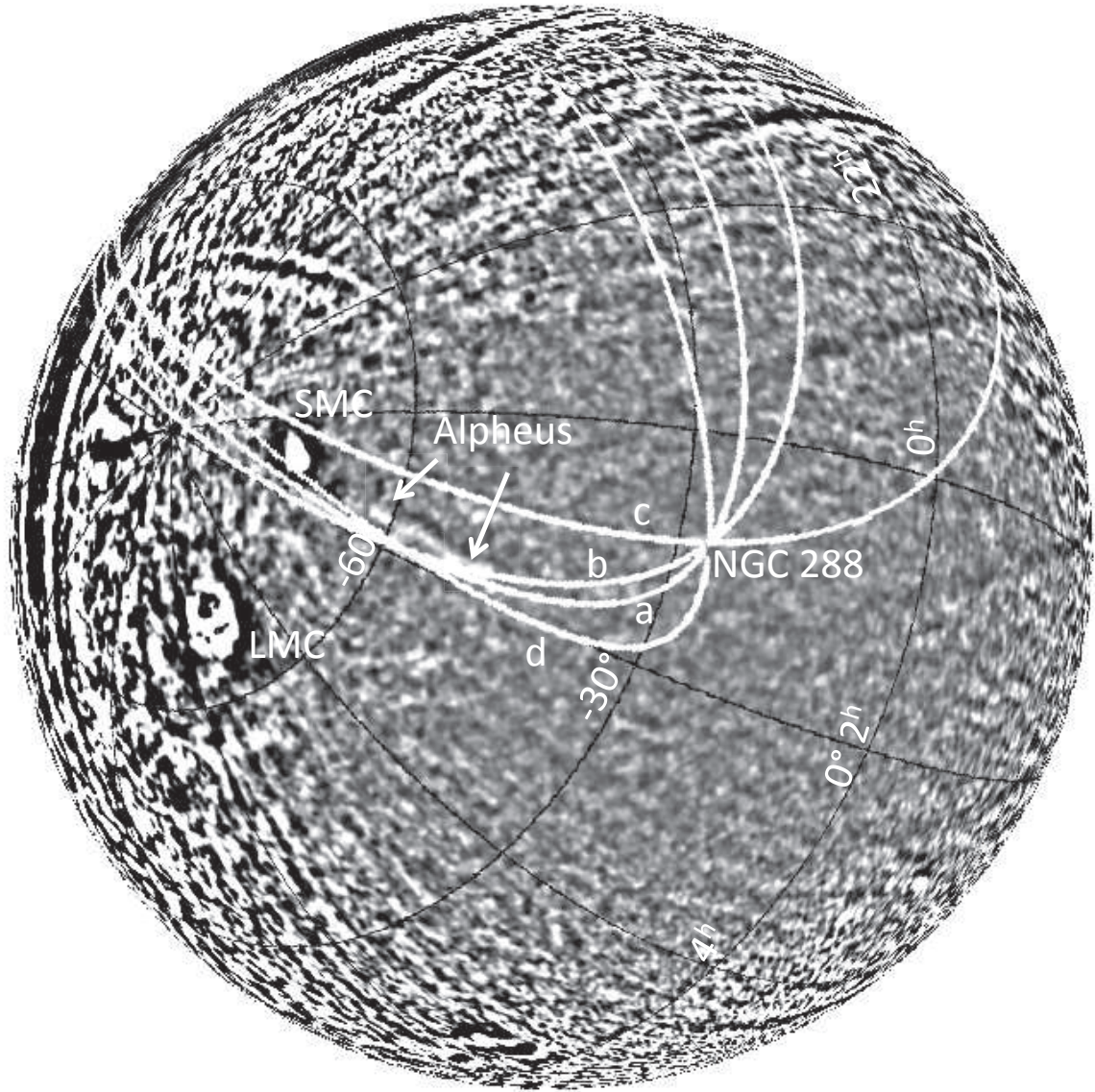}
\caption{Projections of the orbit of NGC 288 using three different
  proper motions measurements (see text). Orbit 'd' was generated by
  both forcing the orbit to pass through NGC 288, and to pass
  2.5\arcdeg~ to the east along the length of Alpheus.}

\end{figure}

\end{document}